\documentclass[12pt]{article}
\usepackage{amsfonts}
\usepackage{latexsym}
\usepackage{amsmath,amssymb}
\usepackage{verbatim}
\usepackage{cite}

\usepackage{setspace}

\usepackage[textheight=9in, textwidth=6.5in, letterpaper]{geometry}
\def\half{{1\over 2}}
\numberwithin{equation}{section}

\def\ip{${\mathcal I}^+$}

\def\g{{\gamma}}

 \def\p{\partial}
 \def\bz{{\bar z}}
 
\def\0{{(0)}}
\def\1{{(1)}}
\def\2{{(2)}}
 \def\cL{{\cal L}}

\def\n{\nabla}
\def\ci{{\mathcal I}}

\def\<{\langle }
\def\>{\rangle }
\def\[{\left[}
\def\]{\right]}

\def\h{{h^A_A}}

\def\o{\omega }
\def\x{{{\cal X}^+}}
\newcommand{\bea}{\begin{eqnarray}}
\newcommand{\eea}{\end{eqnarray}}
\newcommand{\be}{\begin{equation}}
\newcommand{\ee}{\end{equation}}
\newcommand{\ba}{\begin{align}}
\newcommand{\ea}{\end{align}}

\renewcommand{\epsilon}{\varepsilon}

   \makeatletter
  \let\over=\@@over \let\overwithdelims=\@@overwithdelims
  \let\atop=\@@atop \let\atopwithdelims=\@@atopwithdelims
  \let\above=\@@above \let\abovewithdelims=\@@abovewithdelims
\renewcommand\section{\@startsection {section}{1}{\z@}%
                                   {-3.5ex \@plus -1ex \@minus -.2ex}
                                   {2.3ex \@plus.2ex}%
                                   {\normalfont\large\bfseries}}

\renewcommand\subsection{\@startsection{subsection}{2}{\z@}%
                                     {-3.25ex\@plus -1ex \@minus -.2ex}%
                                     {1.5ex \@plus .2ex}%
                                     {\normalfont\bfseries}}

\begin{document}
\begin{titlepage}
\unitlength = 1mm
\ \\
\vskip 3 cm
\begin{center}

{ \LARGE {\textsc{Superrotation Charge and  Supertranslation Hair on Black Holes}}}

\vspace{0.8cm}
Stephen W. Hawking$^\dagger$, Malcolm J. Perry$^\dagger$ and Andrew Strominger$^*$

\vspace{0.8 cm}
\begin{abstract}
It is shown that black hole spacetimes in classical Einstein gravity are characterized by, in addition to
their ADM mass $M$, momentum $\vec P$, angular momentum $\vec J$ and boost charge $\vec K$, an infinite head of supertranslation hair. 
The distinct black holes  are distinguished by classical superrotation charges measured at infinity. 
Solutions with supertranslation hair are diffeomorphic to the Schwarzschild spacetime,  but the diffeomorphisms are part of the BMS subgroup and act  nontrivially on the physical phase space. It is shown that a black hole can be supertranslated by throwing in an asymmetric shock wave. A leading-order Bondi-gauge expression is derived for the linearized horizon supertranslation charge and shown to generate, via the Dirac bracket, supertranslations 
on the linearized phase space of gravitational excitations of the horizon.  The considerations of this paper are largely classical augmented by comments on their implications for the quantum theory. 
 \end{abstract}

\vspace{1.0cm}

\end{center}

\vspace{4 cm}
{\it $^\dagger$DAMTP,
Centre for Mathematical Sciences, University of Cambridge, Cambridge, 
UK}

{\it $^*$Center for the Fundamental Laws of Nature, Harvard University,
Cambridge, MA, USA}

\end{titlepage}

\pagestyle{empty}
\pagestyle{plain}

\def\gzz{\gamma_{z\bz}}
\def\vx{{\vec x}}
\def\p{\partial}
\def\po{$\cal P_O$}
\def\cN{{\cal H}^+ }
\def\N{${\cal H}^+  ~~$}
\def\G{\Gamma}
\def\l{{\ell}}
\def\ch{{\cal H}^+}
\def\Q{{\hat Q}}
\def\T{\hat T}
\def\C{\hat C}
\pagenumbering{arabic}

\tableofcontents

\section{Introduction}
Over the last few years it has been found \cite{Strominger:2013lka,Strominger:2013jfa,He:2014laa,Hyun:2014kfa,Adamo:2014yya,He:2014cra,Campiglia:2015yka,Campiglia:2015qka,Kapec:2015ena,Avery:2015gxa,Campiglia:2015kxa,Avery:2015iix,Lysov:2015jrs,whr} that empty space is a richer place than was previously believed.\footnote{Prescient early work appears in
\cite{ash,Balachandran:2013wsa}.} Even the classical Minkowskian vacuum, far from being a unique, featureless configuration, is infinitely degenerate in all electromagnetic, Yang-Mills and gravitational theories. Information about the vacuum configuration is holographically stored at the asymptotic boundary of spacetime. Different vacua are related by infinite-dimensional asymptotic symmetries which, in the quantum theory, can be infinitesimally described as creating or annihilating soft ($i.e.$ zero-energy) particles such as photons or gravitons.  

The infinity of associated conserved charges constrain every scattering process in asymptotically Minkowskian spacetimes, including those  in which black holes are formed and then evaporate. For each and every conserved charge, the charge on the black hole 
must be reduced (increased) by exactly the amount carried by any emitted (absorbed) particles \cite{Strominger:2014pwa,Pasterski:2015tva,Hawking:2016msc,Avery:2016zce}.  Charge conservation is possible only if black holes themselves carry an infinite number of charges or, equivalently, have an infinite head of `soft hair' \cite{Hawking:2016msc}.  This does not violate the classical no-hair theorems \cite{Chrusciel:2012jk} because the distinct black holes are related by diffeomorphisms, albeit `large' ones which comprise the asymptotic symmetry group and act nontrivially on the classical phase space. 
Soft hair has implications for the information paradox \cite{swh}, since charge conservation enforces quantum correlations between the outgoing Hawking quanta and the soft hair configuration. 

In this paper we undertake a study the properties of the charges arising from infinite-dimensional gravitational symmetries in a weak-coupling expansion. The fundamental definitions of these conserved charges will be given below in terms of simple  boundary integrals near spatial infinity. As usual, integration by parts and the  constraint equations can be used to express these charges as three-dimensional `bulk' integrals and thereby associate distinct contributions to the charge from distinct regions of spacetime such as a black hole.

However, even for the simplest of the conserved charges - the ADM energy - this procedure is in the general case fraught with difficulties associated to the choices of slice and gauge. Quantum fluctuations of the spacetime geometry further diminish the utility of such constructions. 
Nevertheless, in the context of  weak coupling, a perturbative analysis of charge conservation in the bulk can be informative. For example it is possible to show, to first order in the gravitational coupling, that the mass of a black hole always increases by the energy flux of radiation across its horizon. A similar picture should exist for all of the conserved charges. For the infinity of electromagnetic charges, such a picture was obtained in \cite{Hawking:2016msc}. In this paper, while also supplying the reader with some pedagogical background, we continue the program of \cite{Hawking:2016msc} and perturbatively analyze in some detail the infinity of so-called supertranslation and superrotation symmetries. Supertranslation (superrotation) charge conservation equates the total incoming energy at each angle to the total outgoing energy (angular momentum) at the opposing angle \cite{Strominger:2013jfa,Kapec:2014opa}. 

After spelling out our notation in section 2.1, in section 2.2 we reiterate the simple origin of the infinity of conserved charges. We show that the very existence of a well-posed scattering problem in asymptotically Minkowskian general relativity requires a boundary condition which matches certain metric components at the top of $\ci^-$ (past null infinity) to those at the bottom of \ip\ (future null infinity). This immediately implies an infinite number of conserved charges, simply from the equality of all the past and future multipole moments of the matched metric data.  Explicit expressions are given for the supertranslation charges arising from the matching of the Bondi mass aspect, as well as the superrotation charges arising from the matching of the angular momentum aspect. The relationship to previous work on asymptotic behavior and the peeling theorem is briefly discussed in section 2.3. Section 3.1 reviews the derivation of supertranslation symmetry as the action via Dirac brackets of the supertranslation charges
on the physical phase space. Section 3.2 reviews the current status of efforts to similarly associate a symmetry superrotation charge conservation. In section 4 we use the  Bondi gauge to continue supertranslations from the boundary into the bulk of the Schwarzschild geometry. The Bondi-gauge metric of an infinitesimally supertranslated Schwarzschild black hole, $i.e.$ a black hole with supertranslation hair, is derived. In section 5 we show, via an explicit Vaidya-type solution,  how a 
black hole can be physically supertranslated by throwing in an asymmetric null shock wave. 
Supertranslated black holes do not carry supertranslation charge because the group is abelian. However in section 6 we find they can and do carry superrotation charges and an explicit expression is given. This provides a classical diagnostic of supertranslation hair (see also \cite{Flanagan:2015pxa,Compere:2016hzt}). 
Section 7 gives a canonical construction of the generators of  linearized Bondi-gauge supertranslations on the future Schwarzschild horizon $\ch$. Section 7.1 reviews the covariant canonical formalism and symplectic form in gravity. Section 7.2 presents the covariant supertranslation charge $\Q_f^{\ch}$ on the horizon.
In section 7.3, a careful fixing of the residual gauge symmetries in Bondi gauge which (unlike supertranslations) are zero eigenvalues of the pre-symplectic form is performed. The symplectic form is then inverted on the physical phase space to obtain the Dirac bracket. Finally it is shown that the charge $\Q_f^{\ch}$ properly generates horizon supertranslations.

During the course of this work strongly overlapping 
results were independently obtained in \cite{Flanagan:2015pxa,Compere:2016hzt}. Related work has also appeared in 
\cite{Banks:2014iha, Penna:2015gza,susy, Hooft:2016itl, Averin:2016ybl, Barnich:2016lyg, Compere:2016jwb, Dai:2016sls, Campoleoni:2016vsh, Afshar:2016wfy, Sheikh-Jabbari:2016unm, Iofa:2016quw, Kapec:2016aqd, Sheikh-Jabbari:2016lzm, Frassino:2016oom, Eling:2016xlx, Conde:2016csj, Compere:2016gwf, Hotta:2016qtv, Mao:2016pwq, Setare:2016vhy, Zeng:2016epp, Averin:2016hhm, Afshar:2016uax, Mirbabayi:2016axw, Grumiller:2016kcp, Donnay:2016ejv, Betzios:2016yaq, Delgado:2016zxv, Sheikh-Jabbari:2016npa, Carlip:2016lnw, Gasperin:2016dyt, Cai:2016idg, Yan:2016vns, Hopfmuller:2016scf, Epp:1995uc, Brady:1995na, Parikh:1997ma, Reisenberger:2012zq,Parattu:2016trq, Parattu:2015gga, Lehner:2016vdi, Hayward:1993my, Hawking:1996ww, Neiman:2013lxa, Schoutens:1993hu}.
We expect our horizon analysis is closely related to much earlier work \cite{Hotta:2000gx,Koga,Hotta:2002mq,Ashtekar:2004gp} employing different gauges and formalisms. Soft hair appears to be an alternate description of the phenomenon of edge modes as discussed in \cite{Donnelly:2014fua,Donnelly:2015hxa,Harlow:2015lma,
Maldacena:2016upp,Harlow:2016vwg}. A precise characterization of the relation of these edge modes and soft hair would be of great interest. 

We set Newton's constant $G=1$ throughout. 

\section{Supertranslation and superrotation charge conservation}

In this section we review a few salient facts about asymptotically flat spacetimes in classical general relativity (GR) and the newly-discovered infinite number of conserved supertranslation  \cite{Strominger:2013jfa} and superrotation \cite{Cachazo:2014fwa, Kapec:2014opa,Campiglia:2015yka} charges. Moreover, we show that the existence of this infinite number of conserved charges in GR follows simply from the requirement of a well-posed
scattering problem. 
\subsection{Asymptotic expansion}
Near future null infinity (\ip) we use retarded coordinates $ (u,r, \Theta^A)$ and the Bondi gauge, in which 
\be  \label{sx} g_{rA}=g_{rr}=0,~~~~{\rm det}\left({g_{AB} \over r^2}\right)=g(\Theta).\ee 
Surfaces of constant retarded time $u=t-r$
 are null. \ip\ is the surface $r \rightarrow \infty$. $\Theta^A$ are coordinates on the two-sphere and $g(\Theta)$ is a fixed function on the sphere. An asymptotically flat metric has a large $r$-expansion\footnote{See \cite{bt} and \cite{Madler:2016xju} for  recent reviews. We omit here interesting logarithmic terms \cite{Chrusciel:1993hx,Chrusciel:2014lha} of potential relevance in the present context.}
 \bea \label{mt}ds^2=&& -du^2-2dudr+r^2\gamma_{AB}d\Theta^Ad\Theta^B\cr&& \cr&&+{2m\over r}du^2+rC_{AB}d\Theta^Ad\Theta^B +D^BC_{AB}dud\Theta^A \cr&& \cr&&+{1 \over 16r^2}C_{AB}C^{AB}dudr\cr&&\cr&&+{ 1 \over r}\left({4\over 3}N_A+{4u \over 3}\p_Am-{1 \over 8}\p_A(C_{BD}C^{BD})\right)dud\Theta^A\cr&&\cr&&
+{1\over 4}\gamma_{AB}C_{CD}C^{CD}d\Theta^Ad\Theta^B\cr&&\cr &&+\ \ \ldots\ldots\eea
 where indices are raised and lowered using the metric on the unit sphere, $\gamma_{AB}$. The traceless tensor $C_{AB}$, the Bondi mass aspect $m$, and the angular momentum aspect $N_A$ all depend on the \ip\ coordinates $(u,\Theta^A)$ but not $r$. 
 Our definition of $N_A$ \footnote{Our angular momentum aspect $N_A$ can be related to that defined by Barnich and Troessaert \cite{bt}, $N_A^{BT}$ by the following  
$ N_A = N_A^{BT}-D_A m +{1\over 4}C_{AB} D_C C^{BC} + {3\over 16}C^{BC} D_A C_{BC}$.} differs from the conventional ones in two ways. Firstly, by a shift of $u\p_A m$ which has the advantage that, as can be seen from the constraint equations below,  is typically finite for $u\to \pm \infty$. Secondly, there is a shift of $N_A$ by quadratic terms in $C_{AB}$ to obtain a simple relation to the Riemann tensor
 \be \lim_{r\to \infty}r^3R_{Arru}= N_A + u\partial_Am.\ee
The Bondi news  \be\label {bn} N_{AB}=\p_uC_{AB}\ee characterizes the gravitational radiation passing through \ip. 
 
The Cauchy data on \ip\ for the full spacetime metric includes\footnote{Determining if or when these (or the $\ci^-$ counterparts) comprise a $complete$ set of Cauchy data is an outstanding problem in mathematical relativity which we do not address here. Among other issues are  the possibilities of logarithms and further integration functions appearing at higher order in the $1 \over r$ expansion. See for example \cite{Chrusciel:1993hx,Chrusciel:2014lha}.} $C_{AB}, m$ and $N_A$ which are subject to the constraint equations
\bea \label{ct} \p_u m &= &\frac{1}{4} D^AD^BN_{AB}-T_{uu},\cr~~~ T_{uu}&\equiv &{1 \over 8} N_{AB} N^{AB} +4\pi \lim_{r\to \infty}\big[ r^2 T^M_{uu}\big],  \eea
  \bea \label{nct} \p_{u} N_A&=&-{1 \over 4} D^B(D_BD^CC_{CA}-D_AD^CC_{BC})
  +{u }\p_A (T_{uu}-{1\over 4}D^BD^CN_{BC})- T_{uA},\cr T_{u A} &\equiv& 8\pi \lim\limits_{r\rightarrow\infty}[r^2 T^M_{uA}]-{1\over 4 }\p_A(C_{BD}N^{BD})+{1\over 4}D_B(C^{BC}N_{CA})-{1\over 2}C_{AB}D_CN^{BC}.\eea
  Here $T^M_{ab}$ is the matter stress tensor while $T_{ab}$ incorporates corrections from the  stress tensor for linearized gravity waves.
  
The traceless Bondi news $N_{AB}(u,\Theta)$ comprises two unconstrained real functions on \ip\ as expected for the two helicities of the massless graviton. We assume that near the past and future boundaries of \ip, $\ci^+_+$ and $\ci^+_-$, the news falls off faster than $1\over |u|$ and that the angular momentum aspect $N_A$ approaches a finite one-form  on $S^2$. These (and stronger) asymptotic boundary conditions were proven by Christodoulou and Klainerman \cite{ck} to hold in a finite neighborhood of flat space: here we shall consider spacetimes with this asymptotic behavior but do not require them to be near flat space in the deep interior. The news then trivially determines $C_{AB}$ up to an integration 
function by integrating  (\ref{bn}).  We take the integration function to be $C_{AB}|_{\ci^+_-} $.  Finiteness of $N_A|_{\ci^+_-}$ and (\ref{nct}) then imply $C_{AB}|_{\ci^+_-} $ is determined from a single real function $C$ on $\ci^+_-$~\cite{Strominger:2013jfa}:
\be C_{AB}|_{\ci^+_-}=-2D_AD_BC|_{\ci^+_-}+\gamma_{AB}D^2C|_{\ci^+_-}. \ee
Given the news tensor and this initial data at $\ci^+_-$, the constraints may be integrated to give the mass and angular momentum aspects $m$ and $N_A$ everywhere on \ip. Hence the Cauchy data includes 
\be \label{des} \{N_{AB}(u,\Theta), C(\Theta)|_{\ci^+_-}, m(\Theta)|_{\ci^+_-},N_A(\Theta)|_{\ci^+_-}\}.\ee

The Cauchy data (\ref{des}) transforms non-trivially under the BMS+ subgroup\footnote{The `+' signifies the action is on \ip\ rather than $\ci^-$.} \cite{bms} of diffeomorphisms acting near \ip, which includes boosts, rotations and supertranslations (formulae for which are in the next section). As shown in
\cite{bms}, despite being diffeomorphic,  data sets which differ by BMS transformations are physically inequivalent. 
For example they can change the ADM energy or transform a configuration with gravity waves simultaneously emerging at the north and south pole on \ip\ into ones where they appear with an arbitrary relative retarded time delay. Even when the news is zero, BMS+ generically changes the vacuum to an inequivalent one with different values of both $C|_{\ci^+_-}$ and ADM  angular momentum. 
That is, there is an infinite family of inequivalent vacua in GR. 

A similar set of equations apply near $\ci^-$, where we employ advanced Bondi coordinates $(v,r,\Theta^A)$ in which the metric has the asymptotic expansion   
  \bea \label{mtp}ds^2=&& -dv^2+2dvdr+r^2\gamma_{AB}d\Theta^Ad\Theta^B\cr&&\cr&&+{2m\over r}dv^2+rC_{AB}d\Theta^Ad\Theta^B -D^BC_{AB}dvd\Theta^A \cr&&\cr&&-{1\over 16r^2}C_{AB}C^{AB}dvdr\cr&&\cr&&-{ 1 \over r}\left({4\over 3}N_A-{4v\over 3}\p_Am-{1 \over 8}\p_A(C_{BD}C^{BD})\right)dvd\Theta^A \cr&&\cr&& +{1 \over 4}\gamma_{AB}C_{CD}C^{CD}d\Theta^A d\Theta^B \cr&&\cr&&+ \ldots\ldots \eea
  In flat Minkowski space advanced and retarded Bondi coordinates are related by 
  \be\label{pf} (v,r,\Theta^A)=(u+2r,r,P\Theta^A) \ee
  where $P\Theta^A$ is the antipode of $\Theta^A$ on the sphere.\footnote{For standard angular coordinates $\Theta^A\sim (\theta,\phi),~~~P\Theta^A\sim (\pi-\theta,\phi+\pi)$. This coordinate convention is chosen to simplify the past-future matching conditions below.} The analog of the Cauchy data (\ref{des}) for $\ci^-$ is \be \label{de} \{N_{AB}(v,\Theta), C(\Theta)|_{\ci^-_+}, m(\Theta)|_{\ci^-_+},N_A(\Theta)|_{\ci^-_+}\}.\ee  

\subsection{The scattering problem}
The scattering problem in classical general relativity is, roughly speaking, to find the map from Cauchy data on $\ci^-$ to that on $\ci^+$.\footnote{Of course if a black hole is formed we need Cauchy data on \ip$\cup \ch$, where $\ch$ is the future horizon, but this section is mainly concerned with the weak-field problem for which black holes are absent.} Such a map is not even formally determined from the maximal Cauchy development of the $\ci^-$ data (\ref{de}) with the Einstein equation. This determines the data on \ip\  at most up to a BMS+ transformation. A prescription is needed to  attach \ip, choose a BMS+ frame and determine the initial values for integrating $m$ and $N_A$ along \ip\ using the constraints. {{\it Without such a prescription, the scattering problem in GR is not defined.}}  In  \cite{Strominger:2013jfa}, 
it was proposed that the BMS+ frame should be determined by the Lorentz and CPT invariant matching conditions 
\be  \label{cmm}C|_{\ci^+_-}(\Theta)=C(\Theta)|_{\ci^-_+}    ,~~~m(\Theta)|_{\ci^+_-}=m(\Theta)|_{\ci^-_+} ,\ee
and in \cite{Kapec:2014opa} the matching condition for the angular momentum aspect 
\be  \label{nm}N_A(\Theta)|_{\ci^+_-}=N_A(\Theta)|_{\ci^-_+} \ee was proposed. This breaks the combined BMS+$\otimes$BMS$-$ action on \ip\ and  $\ci^-$ down to the diagonal subgroup which preserves these conditions. 
 Noting our convention (\ref{pf}) relating $\Theta^A(\ci^+_-)$ and $\Theta^A(\ci^-_+)$, we see that (\ref{cmm}) and (\ref{nm}) {\it antipodally} equate past and future fields near spatial infinity.  At first sight, this antipodal relation appears rather bizarre. However, we expect that it is the only Lorentz and CPT invariant choice and is implicit in most or all GR computations in asymptotically flat spacetimes. In \cite{He:2014laa} the matching condition (\ref{cmm}) was in fact proven to be implicit to all orders in standard weak field perturbation theory by demonstrating its equivalence to Weinberg's soft graviton theorem \cite{Weinberg}. In \cite{Cachazo:2014fwa,Geyer:2014lca,White:2014qia}  a new subleading soft graviton theorem was proven to all orders using Feynman tree diagrams\footnote{Although this paper largely concerns classical GR, we note that  (\ref{nm}) is possibly deformed by an anomaly at one loop \cite{Bern:2014oka,Cachazo:2014dia,He:2014bga,Bern:2014vva,Broedel:2014fsa,Broedel:2014bza}. Since $some$ matching relation of the form (\ref{nm}) must exist in order for gravitational scattering to be defined, this suggests that these one loop corrections deform rather than eliminate the conserved charges. This is an important open problem. Some recent progress has appeared in \cite{Kapec:2016jld,Cheung:2016iub}. }, and then shown to imply (\ref{nm}) \cite{Kapec:2014opa,Campiglia:2015yka}.  Motivated by this perturbative analysis, we propose  that (\ref{cmm}),(\ref{nm}) are part of the definition of the scattering problem whenever the fields are sufficiently weak near spatial infinity, even if the interior contains a black hole.  
 
The matching conditions~(\ref{cmm}) and (\ref{nm}) immediately imply that an infinite number of charges are conserved in GR scattering.  Two families of charges are defined at $\ci^+_-$ and $\ci^-_+$ by:
\be \label{sdw}Q^+_f={1 \over 4 \pi}\int_{\ci^+_- }d^2\Theta\sqrt{\g}f m,~~~~Q^-_f={1 \over 4 \pi}\int_{\ci^-_+}d^2\Theta\sqrt{\g}f m,\ee
where $f(\Theta)$ is any function on $S^2$. Integrating by parts and using the constraint (\ref{ct}), these can be written as integrals over $\ci^+$ or $\ci^-$ respectively:\footnote{In the presence of massive matter or black holes there are extra contibutions at $\ci^+_+$ and $\ci^-_-$.} \bea \label{sdxw}Q^+_f&=&{1 \over  4\pi}\int_{\ci^+}dud^2\Theta\sqrt{\g}f\left( T_{uu}  - \frac{1}{4} D^AD^BN_{AB}\right),\cr~~~~Q^-_f&=&{1 \over 4 \pi}\int_{\ci^-}dvd^2\Theta\sqrt{\g}f \left( T_{vv}  - \frac{1}{4} D^AD^BN_{AB}\right).\eea
 (\ref{cmm}) implies:
\be\label{dsa} Q^+_f=Q^-_f.\ee 
 The case $f=1$ is just the total energy conservation while the $\l=1$ harmonic $f=Y^1_m$ gives  the well known ADM momentum conservation. The general case (\ref{dsa}) provides an infinite number of new generalizations of these four laws referred to as supertranslation charge conservation \cite{Strominger:2013jfa}. Choosing $f$ to be a delta function, the generalized conservation law equates the net incoming energy flux at each angle (including linear gravitational terms) to the net outgoing energy flux at the opposing angle. The relation of these charges to supertranslation symmetry will be discussed in the next section.

A second infinity of conserved charges can similarly be constructed from an arbitrary vector field $Y^A$ on the sphere.  Using (\ref{nm}) one finds 
 \be \label{asd} Q^+_Y={1 \over 8 \pi}\int_{\ci^+_- }d^2\Theta\sqrt{\g}Y^AN_A={1 \over 8 \pi}\int_{\ci^-_+}d^2\Theta\sqrt{\g}Y^AN_A=Q^-_Y.\ee
 This expresses conservation of superrotation charge. The special cases for which $Y^A$ is one of the 6 global conformal Killing vectors on $S^2$ are conservation of ADM angular momentum and boost charge, sometimes referred to as the BORT (Beig-O'Murchada-Regge-Teitelboim) \cite{bort} center-of-mass. Choosing the vector field to be a delta-function, 
 these new conservation laws equate net in and out angular momentum flux for every angle. 

 The supertranslation and superrotation charges are absolutely conserved in the sense that each gives a number constructed according to (\ref{sdxw}) from incoming classical data on $\ci^-$ that must equal a  number constructed from outgoing data on \ip. This same number can also be constructed  from data on any spacelike slice that ends on $\ci^+_-$ or $\ci^-_+$. This is qualitatively different from $e.g.$ the oft-discussed Bondi mass as a function of  retarded time which is not conserved but rather obeys a conservation law relating its time dependence to energy flux through \ip. 

  The existence of these conserved charges is in principle experimentally verifiable.  Indeed, proposed  tests of the gravitational memory effect, although not initially recognized as such, are tests of supertranslation charge conservation \cite{Strominger:2014pwa}. Superrotation charge conservation may in principle be tested via the gravitational spin memory effect \cite{Pasterski:2015tva}. 
 
 In conclusion, {\it the very existence of a well-posed scattering problem from $\ci^-$ to $\ci^+$ in GR necessitates the existence, for any matching condition, of an infinite number of conserved supertranslation and superrotation charges.} With our matching conditions ~(\ref{cmm}), (\ref{nm}) the explicit expressions for these charges are in (\ref{sdw}), (\ref{asd}).
 
 \subsection{Discussion}
It may seem peculiar that this infinity of exactly conserved charges, which generalize ADM energy and angular momentum,  has gone unnoticed in the more than half a century since the notion of an asymptotically flat spacetime was introduced in \cite{Arnowitt:1959ah}.  Part of the reason for this is that many early studies concentrated on special spacetimes in which the peeling theorem \cite{Sachs:1961zz} applies and Penrose's conformal compactification \cite{Penrose:1964ge} can be utilized. In fact the peeling theorem does {\it not} apply in generic physical settings, see $e.g.$ \cite{dk,md} . A simple example which violates peeling is  a pair of massive bodies 
coming in from infinity with asymptotically constant velocities and no incoming  radiation. 
In this type of situation, however,  peeling can typically be restored by adding incoming radiation in just such a fine-tuned way that the solution is exactly Schwarzschild outside some arbitrarily large but finite radius \cite{Friedrich:1986rb,Habisohn:1989sy,Corvino:2003sp}. 
This procedure fine-tunes all of the nontrivial supertranslation and superrotation charges to zero, rendering the conservation laws rather trivial. 
 It was a commonly held expectation that, in the generic physical case, the singularity structure near spatial infinity is too uncontrolled to admit well-defined conserved charges of the type described here. If correct, this would suggest that there are no physical contexts in which an infinite number of non-trivial and well-defined conserved charges can exist.  This all changed relatively recently starting from the work of Christodoulou and Klainerman \cite{ck}, who showed\footnote{  The key result of \cite{ck} relevant for our purposes is that the Bondi news falls off at least as fast as $1 \over |u|^{3/2}$ (or $1 \over |v|^{3/2}$ ) near the boundaries of $\ci$. This is much faster than required for finiteness of the total radiated energy, and in particular implies that $C_{AB}$ is finite and well defined at the boundaries of $\ci$, enabling the fundamental identification (\ref{cmm}). If the news decayed only as $1 \over |u|$, $C_{AB}$ would diverge and the scattering problem would be ill-posed.} that generic spacetimes in a finite neighborhood of flat space lie precisely in the sweet spot where it is possible to define \cite{Strominger:2013jfa,Kapec:2014opa,Campiglia:2015yka} an infinite number of finite, generically non-zero and conserved  supertranslation and superrotation charges. In this paper we consider a larger family  of spacetimes whose asymptotics lie in the same sweet spot and have the conserved charges, but are not necessarily in a small neighborhood of flat space and may contain black holes in the interior.  
 \section{Asymptotic  symmetries}
 
It is typically the case that conserved charges imply symmetries.  In judicious circumstances, a physical phase space $\Gamma$ can be 
constructed by imposing suitable constraints and gauge conditions. Dirac's procedure is then applied to give the Dirac bracket 
$\{~,~\}$. One then defines the infinitesimal symmetry associated to a conserved charge $Q$ on the fields $\Phi$ 
by 
\be \delta\Phi=\{Q,\Phi\}.\ee
In practice, many subtleties may arise in implementing this procedure including the identification of proper boundary conditions and zero modes. 
As reviewed in this section, the program has been completed for supertranslations 
but remains underway for superrotations.
\subsection{Supertranslations}Dirac brackets involving $C_{AB}$ (including its zero modes) were constructed in \cite{He:2014laa}.\footnote{This refined the results of \cite{Palmer:1978zz, ash, Crnkovic:1986ex,wz} by a careful treatment of zero modes, including an imposition as physical constraints of the vanishing of the Weyl tensor and the Bondi news at the boundaries of $\ci$. } Commutators of the supertranslation charge $Q^+_f$ in 
(\ref{sdw}) were then shown to obey  
\be \{Q^+_f, C_{AB}\}=f\p_u C_{AB}-2D_AD_Bf+\gamma_{AB}D^2f,\ee
\be \{Q^+_f, C|_{\ci^+_-}\}=f.\ee
This is easily recognized as the action on $C_{AB}$ of the BMS+ supertranslations \cite{bms} which are diffeomorphisms generated by the vector field 
\be\label{ssx} \zeta_f=f\p_u-{1 \over r}D^A f \p_A +\half D^2f \p_r +...\ee
Here the subleading ${1 \over r}$ corrections required to preserve Bondi gauge depend on the metric and $D^2\equiv \g^{AB}D_AD_B$.  The full BMS+ group is a semidirect product of supertranslations with the Lorentz group. 

The fact that the symmetry generated by $Q_f$ is a subgroup of a known symmetry (diffeomorphisms) of the standard presentation of the theory is a beautiful feature of this example. It is not obvious or a priori guaranteed. Indeed there are a number of examples  ($e.g.$ \cite{susy}) where this is not the case.

Interestingly, the vacuum solution $C_{AB}=0$ on \ip\ is not invariant under supertranslations. In other words, supertranslation symmetry is spontaneously broken. 
There are an infinite number of degenerate classical vacua  labelled by the function $C|_{\ci^+_-}$, each of which is preserved by a different Poincare subgroup of BMS+. These vacua have different ADM angular momenta.  This is consistent with the existence of  vacuum solutions with nonzero angular momentum \cite{Chen:2014uma}. This is sometimes referred to as the `problem of angular momentum' in GR. However properly understood it is a beautiful feature indicating a rich vacuum structure, not a problem! 

\subsection{Superrotations}

It is natural to expect that superrotation charges canonically generate the antipodally-identified Virasoro-like symmetry 
presented in \cite{bt, Kapec:2014opa} whose global $SL(2,C)$ subgroup is the Lorentz group. We think this is likely in some sense the case. However 
superrotation symmetry is more subtle than its supertranslation analog and the construction has not been completed. The difficulty can be seen in a naive application of the  Dirac brackets of 
\cite{He:2014laa}  which yield
\be \{Q^+_Y, N_{AB}\}=\cL_Y N_{AB}-D_AD_B D_CY^C+\frac{1}{2}\gamma_{AB}D^2D_CY^C.\ee
Apparently $Q^+_Y$ does not preserve the condition that $N_{AB}$ vanish at the boundaries of $\ci^+$: $i.e.$ it does not map the phase space considered in \cite{He:2014laa} into itself. Quantum mechanically, the action of $Q^+_Y$ will produce a state outside the Hilbert space studied in \cite{He:2014laa}.  A larger phase space and associated bracket is needed, but has not yet been found. Indeed recent work \cite{Strominger:2016wns } building on \cite{penrose_book} has shown  have shown that superrotations can create strings which pierce $\ci$ and destroy asymptotic flatness, suggesting the requisite phase space is the one considered in \cite{Ashtekar:1981ar}. Other very interesting recent works have suggested that superrotations can be understood in terms of diffeomorphisms which violate standard falloff conditions \cite{Campiglia:2016jdj,Campiglia:2016hvg,Conde:2016csj,Strominger:2016wns}.  An important issue for the quantum theory is the appearance of one loop corrections \cite{Bern:2014oka} which depend on the  order of soft limits \cite{Cachazo:2014dia}. These and other important issues are beyond the scope of this paper (although in section 7.3 we will show that non-holomorphic superrotations preserve Bondi gauge). Early discussions of superrotation symmetry can be found in \cite{penrose_book, deBoer:2003vf,banks}, and more recent ones in \cite{bt, Cachazo:2014fwa,Kapec:2014opa,Campiglia:2015yka,Flanagan:2015pxa,Donnay:2015abr,   Kapec:2016jld,Cheung:2016iub}. 

In this paper we will not use the superrotation $symmetry$ per se - only the finite and conserved superrotation $charge$ given by (\ref{asd}), and defer the above interesting issues to future work.

\section{Schwarzschild supertranslations}

In this section we will describe the infinitesimal supertranslation of the Schwarzschild black hole $i.e.$ a black hole with linearized supertranslation hair.
This specializes more general formulae which can be found in \cite{bt}.
This type of soft hair appears to be an alternate description of the edge modes as discussed in \cite{Donnelly:2014fua,Donnelly:2015hxa,Harlow:2015lma,
Maldacena:2016upp,Harlow:2016vwg}.

The extension of an asymptotic gauge symmetry into the interior is gauge dependent. 
In a general time dependent situation, there is unlikely to be a useful or canonical choice of gauge. 
Quantum fluctuations further diminish the utility of specific choices.  In quantum gravity in asymptotic Minkowski space, we expect the only fully well-defined observables are supported at the boundary at infinity.

It is nevertheless sometimes possible, armed with a gauge choice,  to define interior quantities such as local gravitational 
energy densities at first non-trivial order in perturbation theory around Schwarzschild. 
This is sometimes useful in developing a picture and intuition for the behavior of the spacetime away from its boundary. For example one may show at leading order in perturbation theory that, at both the classical and quantum level, the total energy comprised of linearized perturbations plus the mass of the black hole itself is conserved. Moreover, this perturbative conservation law is the linearization of an exact, nonperturbative conservation law, which can only be exactly phrased in terms of asymptotic quantities.  It is in this spirit that we study the linearized action of supertranslations in Schwarzschild.

In advanced Bondi coordinates the Schwarzschild metric is  \begin{equation}
  \label{eq:2}
  ds^2=-V dv^2+2 dvdr+r^2
\g_{AB}d\Theta^Ad\Theta^B\,,~~~~V\equiv1-{2M \over r}.
\end{equation}
We wish to find the BMS$-$ supertranslations $\zeta$ which preserve Bondi gauge (\ref{sx}) and the standard metric component falloffs at large $r$ while having bounded components in a local orthonormal frame at large $r$. The last condition eliminates all superrotations, including boosts and rotations. 
The former conditions require, for Schwarzschild
\bea \label{gone} \cL_\zeta g_{rA}
=\p_A \zeta^v+g_{AB}\p_r\zeta^B =0,\eea
\vskip-37pt
\bea \label{gtwo} \cL_\zeta g_{rr}=2\p_r \zeta^v=0,\eea
\vskip-20pt
\be \label{gthree}  {r \over 2} g^{AB}\cL_\zeta g_{AB}=rD_A\zeta^A+2\zeta^r=0.\ee
The general  solution to this consistent with Bondi gauge and asymptotic falloffs is\footnote{The general solution without restricted falloffs is given in section 7 below.}
\be\label{ssy} \zeta_f=f\p_v+{1 \over r}D^A f \p_A -\half D^2f \p_r 
,~~~ \p_r f=\p_vf=0.\ee 
This extends the asymptotic expansion of the supertranslations on $\ci^-$
to the entire region covered by the advanced coordinates. This includes 
$\ci^-$ and $\ch$ but not \ip.
These act on the Schwarzschild metric as  
\bea \label{ssty} \cL_fg_{vv} &=&{MD^2f \over r^2},\cr \cr \cL_f g_{AB}&=&2rD_AD_Bf 
-r\gamma_{AB}D^2f,\cr\cr
\cL_f g_{Av}&=&-D_A(Vf+\half D^2f).\eea
Adding this to  (\ref{eq:2}) gives the infinitesimally supertranslated Schwarzschild geometry:\footnote{ It may be possible to find the finitely supertranslated geometry. This was accomplished at null infinity in \cite{Barnich:2016lyg} and related finite problems were solved in \cite{penrose_book,Barrabes:1991ng,
Blau:2015nee,Compere:2016hzt,Blau:2016juv}. However in this paper our attention is restricted to the linearized theory. } \bea\label{sf} ds^2&&=-(V -{MD^2f \over r^2})dv^2+2 dvdr- dvd\Theta^AD_A(2Vf+D^2f) \cr\cr &&~~~~~~+
(r^2\g_{AB}+2rD_AD_Bf-r\g_{AB}D^2f)d\Theta^Ad\Theta^B .\eea
The event horizon is at $r=2M+\half D^2f$.  This describes a black hole with linearized supertranslation hair. 
 
\section{Implanting supertranslation hair}

In the previous sections we described a supertranslated eternal Schwarzschild black hole. In order to be certain such objects really exist, in this section 
we describe how one physically makes such a hairy black hole. 

First we show how to add supertranslation hair to bald eternal Schwarzschild, and then generalize to a black hole formed from the vacuum. 
At advanced time $v_0$ in Schwarzschild we send in a  linearized shock wave with energy momentum density   
\be \T_{vv}={\mu+\T(\Theta)\over 4\pi  r^2}\delta(v-v_0) \ee near $\ci^-$. We wish to solve for the linearized metric in such a way that the solution is diffeomorphic to Schwarzschild both before and after the shock wave. Stress energy conservation $\nabla_a\T^{ab}=0 $
then mandates subleading in ${1 \over r}$ corrections 
to the stress tensor for shock waves which are not spherically symmetric. These take the form\footnote{We are grateful to Alex Lupsasca, Monica Pate and Prahar Mitra for help with this solution.}
\be \label{dti} \T_{vv}=\left({\mu+\T\over 4\pi  r^2}+{\T^\1 \over 4\pi  r^3}\right)\delta (v-v_0),~~~\T_{vA}= {\T_A \over 4\pi  r^2} \delta(v-v_0).\ee where $\T(\Theta)$ has only $\l>1$ components\footnote{The $\l=0$ component is represented by $\mu$. The $\l=1$ component, which would  add $ADM$ momentum to the black hole, is eliminated to simplify the discussion. } so that  and 
$\T^\1$ and $\T_A$  are functions of $x$ determined by 
\be\label{der} (D^2+2)\T^\1 =-6M\T,~~~D^A\T_A=\T^\1.\ee
The solutions are conveniently expressed by introducing the Green function solving 
\be {\sqrt{\g} \over 4}D^2(D^2+2)G(\Theta;\Theta')= {\delta^2(\Delta\Theta) },\ee
namely \cite{Strominger:2014pwa}
\be G(\Theta;\Theta')={1\over \pi}\sin^2{\Delta \Theta \over 2}\log \sin^2{\Delta \Theta \over 2} ,\ee
where $\Delta \Theta$ is the angle on the sphere between $\Theta$ and $\Theta'$. Further defining 
\be \C(\Theta)\equiv \int d^2\Theta^\prime G(\Theta,\Theta')\T(\Theta'),\ee
(\ref{der}) has the solution 
\bea \T&=& {1 \over 4}D^2(D^2+2)\C, \cr\cr \T^\1&=&-{3M\over 2}D^2\C,\cr\cr
\T_A&=&-{3M \over 2}\p_A\C.\eea
Equivalently
\begin{equation}\label{src}
\begin{split}
\T_{vv} &= \frac{1}{4\pi r^2} \left[   \mu + \frac{1}{4} D^2 \left( D^2 + 2 \right) \C - \frac{3 M }{ 2r } D^2 \C  \right] \delta \left( v - v_0 \right)  ~, \\ \\
\T_{vA} &= - \frac{3 M }{ 8\pi r^2 } D_A \C\delta \left( v - v_0 \right) ~. 
\end{split}
\end{equation}
The leading large-$r$ constraint equation  on $\ci^-$ may then be written \be \label{cta} \p_v m = \frac{1}{4} D^AD^BN_{AB}
+(\mu+\T(\Theta))\delta (v-v_0). \ee
This equation constrains, but does not fully determine, the mass aspect $m$ and $C_{AB}$.
We wish to solve it in such a way that $\p_A m=0$ everywhere. Integrating over the sphere this implies
\be \label{mb} m=M +\mu \theta(v-v_0).\ee
It then follows from (\ref{ct}) that
\be D_AD_BC^{AB} =-4\T(\Theta)\theta(v-v_0).\ee
The unique solution to this is 
\be C_{AB}=-2\theta(v-v_0)(D_AD_B\C-\frac{1}{2}\gamma_{AB}D^2\C).\ee
One may verify that  \bea\label{mp}  h_{vv}&=& \theta(v-v_0) (\frac{2\mu}{r}  - {MD^2\C \over r^2}), \cr\cr h_{AB}&=&-2r\theta(v-v_0)(D_AD_B\C -\frac{1}{2}\gamma_{AB} D^2\C),\cr\cr 
h_{vA}&=& \theta(v-v_0)\p_A( 1-{2M \over r}+\half D^2)\C,\eea
solves the linearized Einstein equation with source (\ref{src}) for all $r$ and hence are the complete linearized metric perturbations. Comparing with the formula (\ref{ssty}) for a supertranslation of Schwarzschild one  finds (\ref{mp}) can be written 
\be h_{ab}=\theta(v-v_0)\left(\cL_{f=-\C}g_{ab}+{2\mu \over r}\delta_a^v\delta_b^v\right).\ee
Hence the shock wave is a domain wall interpolating between two 
BMS inequivalent Schwarzschild vacua, whose mass parameters differ by $\mu$. 

The shock wave induces a shift in the transverse components of the metric perturbation on the horizon.  Integrating  over a null generator of the horizon
\be \label{svx} \Delta h_{AB}(r=2M,v,\Theta)=\int dv \p_vh_{AB}(r=2M,v,\Theta)=-4M(D_AD_B\hat C (\Theta)-\frac{1}{2}\gamma_{AB}D^2\hat C).\ee
At the quantum level, the expectation value of the metric perturbation in the semiclassical state produced by the shock wave must have the profile (\ref{svx}). That is, it must be close to a coherent state of soft gravitons. 
According to Weinberg's theorem, soft gravitons at \ip\  are excited whenever energy crosses \ip\  with an
  $\l>1$ angular momentum profile. 
Similarly, soft gravitons at $\ch$ are excited whenever energy is thrown into the black hole with an 
  $\l>1$ angular momentum profile. It would be interesting to see this diagrammatically 
in perturbation theory around Schwarzschild from the emergence of a pole in the  soft limit of gravitons falling into the black hole. It may also be possible to use (\ref{svx}) to define a `black hole memory effect' which can be measured by observers hovering just outside the horizon.  

It is trivial to generalize this construction to a black hole formed from the vacuum via a Vaidya shock wave at $v=v_S$.  One simply replaces the mass aspect appearing in (\ref{mb}) by
\be m=M\theta(v-v_S)+\mu \theta(
v-v_0).\ee
Hence hairy black holes can be classically produced from the vacuum. In the next section we see how they are classically distinguished by their superrotation charges. 

\section{Classical superrotation charges of supertranslation hair}

Supertranslating a black hole does not add supertranslation charges to the black hole, just as an ordinary translation of a black hole does not add momentum.  This follows from the fact that  the supertranslation group is abelian, and may also be seen directly by evaluating the charge expressions of the previous section. However, as supertranslations and superrotations do not commute, a supertranslated black hole can and does carry superrotation charges, already at the classical level. In this section we work out these charges for linearly supertranslated Schwarzschild.

From (\ref{asd}) the conserved superrotation charges are 
 \be \Q^-_Y =\frac{1}{8\pi } \int_{\ci^-_+} d^2\Theta\sqrt{\g}Y^A N_A,\ee
where $Y^A$ is any smooth vector field on the sphere. 
We are interested in the differential  superrotation charges carried by an infinitesimally supertranslated Schwarzschild black hole of the type considered in the previous section. As seen from (\ref{ssty}), under a supertranslation $\delta_fg_{ab}=\cL_f g_{ab}$ of Schwarzschild 
 \be \delta _{f} N_A=-3 M \p_A f .\ee
 It follows immediately  that 
\bea \label{dsk} \Q^-_Y(g,h=\delta _fg)=-{3 \over 8 \pi}\int_{\ci^-_+} d^2\Theta\sqrt{\g}  Y^A M \p_Af. \eea
This is nonzero for a generic vector field $Y^A$ and supertranslation $f$. An infinite number of superrotation charges can be independently added to the black hole by different choices of $f$.  Hence the superrotation charges classically distinguish differently supertranslated black holes. 
Classical black holes sport an infinite head of \lq\lq supertranslation hair"  which is rearranged essentially every time something is thrown into it.\footnote{ Similar observations were made in \cite{Flanagan:2015pxa} and in the context of the membrane paradigm in \cite{Penna:2015gza,Eling:2016xlx}.}

The most easily measurable quantity is the difference in superrotation charges before and after the supertranslation hair  implant. This is because the definition of superrotation charges (including angular momentum) is ambiguous up to conjugation by supertranslations. In particular, in this example, we could conjugate the superrotation charges by $f$ so that  all vanish post-implant. However, the pre-implant charges then become nonzero while  the difference of pre-post implant charges is unaffected. This is analogous to  the standard gravitational memory effect at \ip\ which also measures differences in supertranslation frames.

However one should not conclude from this that only the charge difference is physical, any more than one should conclude that only black hole energy or momentum differences  (a special case) are physical. Measurement of absolute (rather than relative) energy, momentum, angular momentum or any of the superrotation charges is also possible but requires specification of an asymptotic Poincare frame.  In the physical phase space, two black hole spacetimes which differ by any element of BMS correspond to different points. For the case of boosts, the two spacetimes have different energy. For supertranslations they are energetically degenerate, but carry different superrotation charges (including angular momentum) and are still physically distinct points. In the quantum theory, the corresponding states are orthogonal and can be superposed. An important difference between boosts and supertranslations is that the latter act nontrivially on all the zero-energy vacua  as well,\footnote{Any given  boost element of BMS acts nontrivially on a generic vacuum, but  every vacuum is preserved by some Poincare subgroup of BMS. There is no preferred Poincare subgroup \cite{bms}.} imparting superrotation charges at quadratic order \cite{Barnich:2016lyg}. Hence the phase space of asymptotically flat geometries with nonzero energy and 4 Killing vectors is not a simple product of vacuum and  black hole phase spaces.\footnote{As we shall see in the next section, the linearized  supertranslation charge around  a black
hole geometry naturally decomposes into the sum of a horizon term and a $\ci$ term, which are not separately conserved in the general nonlinear context. It may be interesting to consider the phase space action of only one term, but such configurations will generically not be static or have Killing symmetries.}

The formula (\ref{dsk}) of superrotation charges requires only the asymptotic behavior of the black hole, and would in a sense pertain to essentially any configuration with the same mass.  To understand this,  let us suppose we sent the supertranslating shock wave into a star or a collection of stars instead of into a black hole. The wave will excite and rearrange the interior structure of the star and, in the case of multiple stars, their relative motions. Generically gravitational radiation will carry some, but not all, of the superrotation charge back out to infinity, while some will be retained by the star(s). It is unsurprising that a star, or a collection of stars, which has many internal degrees of freedom and possible interior states, can carry many superrotation charges. There is no no-hair theorem for a star.  Now consider instead replacing the black hole by a massive stable `hairless' elementary particle with no internal degrees of freedom. Such an object cannot carry arbitrary superrotation charges: the pre- and post- superrotation charges are generically the same (except for the $\l=1$ component). To leading order, the supertranslating shock wave will simply be reflected through the origin and scatter back up to future null infinity. The elementary particle has no mechanism to absorb  all the superrotation charges.  The outgoing wave will cancel the superrotation charges induced by the ingoing wave and, in the far future, the superrotation charges will revert to their initial incoming values.

So we see that in this sense black holes act more like a complex star with many internal degrees of freedom than
a massive elementary particle. The observer at infinity can confirm this by sending in shock waves and watching what comes out. 

At the same time, we note that the exact definition of supertranslation hair in the nonlinear theory given here relies on the existence of an asymptotically flat spacetime boundary and so is not fully intrinsic to the black hole. 
Although it may be possible, we do not know how to canonically associate supertranslation hair to a classical stationary black hole in AdS (this is likely related to the discussion in \cite{Harlow:2015lma}), while a star in AdS clearly retains many internal degrees of freedom. The next section sheds some light on this issue by giving  an intrinsic definition of the horizon contribution to the supertranslation charge to linear order around Schwarzschild. 

\section{Horizon charges}

In the absence of eternal black holes or massive fields, the linearized supertranslation charges $\Q^+_f$ can be written as volume integrals of local operators over $\ci^+$, as explicitly demonstrated in \cite{He:2014laa}. However for Schwarzschild this is clearly impossible, as \ip\ is not a Cauchy surface. Rather, in the absence of massive fields, 
$\ci^+\cup\ch$ is a Cauchy surface. Hence one expects  a relation of the form \be \label{sro} \Q^+_f=\Q_f^{\ci^+}+\Q_f^{\ch}.\ee
The precise form of the horizon contribution $\Q_f^{\ch}$ will depend on the coordinate choice used to extend the supertranslations in from the boundary to the horizon. Here we use Bondi coordinates for this purpose.  We  gauge fix linearized metric fluctuations of the horizon to obtain a physical horizon phase space $\G_{\ch}$. The symplectic form is then constructed and inverted to obtain the Dirac bracket.  An expression for  $\Q_f^{\ch}$ is derived and shown to generate supertranslations on $\G_{\ch}$.  The construction requires that $\Q_f^{\ch}$ and the soft graviton modes, which are nonvanishing on the boundaries $\ch_\pm$ of $\ch$, be incorporated as symplectic partners within $\G_{\ch}$.

Our construction of course makes sense only in (leading order) perturbation theory: in the general case the classical horizon is defined only nonlocally and, even worse, in the quantum case it evaporates. It is doubtful that in the presence of interactions a clean separation can be made between the two terms on the right hand side of (\ref{sro}).  Nevertheless  we hope it may prove useful in developing intuition for the effects of supertranslation charge conservation on black hole dynamics. 

This section relies heavily on general formulae  from the literature \cite{Ashtekar:1981bq,Crnkovic:1986ex,Zuckerman:1989cx,Lee:1990nz,wz,Barnich:2001jy,Avery:2015rga} on the covariant canonical formalism and symplectic forms in gravity.  

\subsection{Symplectic forms and linearized charges}

We expand in variations  $\delta g_{ab} = h_{ab}$ around a fixed background metric $g_{ab}$ which solves the Einstein equation.
The variations are taken to obey the linearized vacuum Einstein equation 
\be \label{eom}\Box h_{ab} +2R_{acbd}h^{cd}-2R_{c(a}h^{~c}_{b)} 
- \n_a\n_ch^c{}_b - \n_b\n_ch_a{}^c  +\n_a\n_b h = 0. \ee
and so are tangent to the phase space. The general results of \cite{Lee:1990nz} give the pre-symplectic structure as
\be\label{sform} \omega(h,h^\prime) = \int_{\Sigma} *J(h,h^\prime), \ee
where the surface $\Sigma$ is a Cauchy surface in the spacetime and $J $ is the pre-symplectic one-form. Explicitly, 
\be \label{smf} \begin{split} J= \frac{\sqrt{-g}}{16\pi } \Biggl[ \frac{1}{2}h \n^b h^{\prime}{}_{ab}
        -\frac{1}{2} h\n_a h^{\prime}
        &+\frac{1}{2} h_{ab} \n^b h^{\prime}
        -h^{bc} \n_b h^{\prime}_{ac}
        +\frac{1}{2}h^{bc}\n_ah^{\prime}{}_{bc}  \\
        - &( h \leftrightarrow h^{\prime} ) \Biggr]~~dx^a. \end{split} \ee
It may be shown that (\ref{eom}) implies
        \be d*J=0.\ee 
Hence  $\omega(h,h^\prime)$ is a functional of the background metric and the two on-shell
variations $h$ and $h^\prime$.  It is invariant under deformations of the surface $\Sigma$ which leave the boundary $\p\Sigma$ fixed.

We are particularly interested in the case where one of the variations - say $h'$ - is pure gauge
$i. e.$ $h'_{ab} = \n_a\zeta_b + \n_b\zeta_a\equiv h^\zeta_{ab}$. $J$ then becomes coexact 
\be *J={1 \over 16 \pi}d*F ,\ee 
where  $F_{ab}$ is given by 
\be \label{dcz}\begin{split} F_{ab} = &\frac{1}{2}(\n_a\zeta_b - \n_b\zeta_a)h
+(\n_ah^c{}_b - \n_bh^c{}_a)\zeta_c
+(\n_c\zeta_a\ h^c{}_b  - \n_c\zeta_b\ h^c{}_a) \\&
-(\n_c h^c{}_b\ \zeta_a - \n_ch^c{}_a\ \zeta_b)
-(\n_ah\ \zeta_b - \n_bh\ \zeta_a). \end{split} \ee
One then has   \cite{Crnkovic:1986ex,Zuckerman:1989cx,Lee:1990nz,wz,Barnich:2001jy,Avery:2015rga} \be  \label{ccn}\o(h,h^\zeta)=-{1 \over 16 \pi }\int_{\p\Sigma} *F \equiv \Q^{\p\Sigma}_\zeta.\ee 
When $\Sigma$ is a Cauchy surface, and this is nonzero, it indicates that the diffeomorphism $\zeta$ acts non-trivially on the physical phase space. 
$\Q^{\p\Sigma}_\zeta$ is a conserved charge associated to the surface $\p\Sigma$ and diffeomorphism $\zeta$ in the sense that it does not depend on the choice of an interior surface 
$\Sigma$. 
We use the hat notation $\Q$ to emphasize that this is the linearized difference in the charge between the geometries $g$ and $g+h$. We will not in this paper consider the  integration to finite charges.  

In general, the pre-symplectic form $\omega$ has zero modes and cannot be inverted. These are eliminated by imposition of the constraints and a judicious choice of gauge. Once this has been accomplished, the restriction of the pre-symplectic form to the physical subspace is the symplectic form. Since all of the zero modes have now been eliminated, it may be inverted to find the Dirac bracket. Once we have done this,  we may decompose a tangent vector in this physical phase space by 
$h_{ab}=\sum_ih^ih_{iab}$ where the tangent index $i$ runs over the (infinite) dimension of the phase space. Defining the inverse of $\omega$  by 
\be \omega^{ij}\o_{jk}=\delta^i_k,\ee
the Dirac bracket is 
\be \{h^i,h^j\}=\omega^{ij}.\ee
Using 
\be \Q^{\p\Sigma}_\zeta = (h^\zeta)^i\omega_{ij}h^j,\ee
it follows immediately that 
\be \{ \Q^{\p\Sigma}_\zeta,h^i\} = (h^\zeta)^i.\ee
In other words,  $\Q^{\p\Sigma}_\zeta$ infinitesimally generates the action of the symmetry $\zeta$ on the physical phase space via the Dirac bracket. 

\subsection{Schwarzschild charges}

We work in the  Bondi gauge (\ref{sx}) for which
\be \label{sxz} h_{rr}=h_{rA}=\g^{AB} h_{AB}=0, \ee
and are  interested in the case $\p\Sigma$ is an $S^2$ of constant $r$ and $v$. One then has 
 \be \label{cdf}\Q^{\p\Sigma}_\zeta=-{1 \over 16 \pi}\int_{\p\Sigma} d^2\Theta \sqrt{\g}r^2 F_{rv},\ee 
where for Schwarzschild the general expression (\ref{dcz}) reduces to 
\bea \label{frv} F_{rv}&=&\zeta^A(\p_r h_{Av}-{2\over r}h_{Av})+\zeta^v(-{1 \over r^2}D^Ah_{Av}-{2 \over r}h_{vv} ) +\p_r \zeta^v h_{vv} +{1 \over r^2}
D^A\zeta^vh_{vA}\cr
&&~~~+\p_r\zeta^vVh_{vr}- \zeta^v{4V \over r}h_{vr}
+\zeta^r{2 \over r}h_{vr}. \eea
For $\zeta=\zeta_f$ a supertranslation as given in (\ref{ssy}), after discarding total derivatives on $S^2$, we find
\bea \label{fdf}F_{rv}={1 \over r}D^Af\p_r h_{Av}-f\left({2 \over r}h_{vv}+{4V \over r}h_{vr}\right)-D^2f
{1 \over r}h_{vr}. \eea
Given the large $r$ asymptotics, only the  $h_{vv}$ term survives for $r\to \infty$ and one finds 
\be  \Q_{\zeta_f}^{\ci^-_+}={1 \over 4 \pi}\int_{\ci^-_+}d^2\Theta\sqrt{\g} f\hat m,\ee
which is the standard expression for the incoming linearized supertranslation charge $\Q^-_f$. The hat on $\hat m$ denotes that it is the deviation of the Bondi mass aspect from the 
background around which we are expanding. 
Supertranslation charge conservation equates this to the outgoing charge
\be  \Q_f^{\ci^-_+}=\Q^{\ci^+_-}_f={1 \over 4 \pi}\int_{\ci^+_-}d^2\Theta\sqrt{\g} f\hat m.\ee

Assuming there are no massive particles\footnote{The additional boundary terms at $\ci^-_-$ from massive fields are given in \cite{Campiglia:2015kxa}.} or black holes,  after imposing constraints and fixing to the Bondi gauge,
the symplectic form (\ref{sform}) was 
inverted in \cite{He:2014laa} to obtain the Dirac bracket $\{ ~,~\}$. Using the constraints to rewrite the charge as a $\ci^-$ integral \be \Q_f^{\ci^-}=\Q_f^{\ci^-_+}={1 \over 4 \pi}\int_{\ci^-}d^2\Theta dv \sqrt{\g} f(T_{vv}-{1 \over 4}D^AD^BN_{AB}).\ee
It was then shown that on $\ci^-$
\be \{ \Q_f^-,~\}=\delta   _f.\ee
That is,  in this special case the supertranslation charge indeed generates supertranslations via the Dirac bracket.

We would like to preform a similar construction of the charge in the presence of a black hole. 
Let $\x$ be a hypersurface extending from $\ci^+_-$ to $\ch_+$, so that $\x\cup\ch$ is a Cauchy surface for the portion of the spacetime exterior to the black hole. Then\footnote{For eternal Schwarzschild, which has a past horizon, charge conservation will require a matching condition relating data on $\ch_-$ to that on ${\cal H}^-_+$.}
\be \Q^-_f=\Q^\x_f+\Q_f^{\ch}.\ee
$\Q_f^{\ch}$ is the contribution to the full supertranslation charge coming from the horizon: $i.e.$ the black hole supertranslation charge. It is the difference between two boundary terms
\bea  \Q_f^{\ch }&=&{M\over 8\pi }\int d^2\Theta\sqrt{\g} f\left[ D^A\p_r h_{Av}+2h_{vv}+D^2h_{vr}\right]^{\ch_+}_{\ch_-} .\eea
We wish to study its properties and demonstrate that it generates horizon supertranslations. 

$\Q_f^{\ch}$ can be written as a horizon integral by integrating by parts and using the constraints. The linearized constraints on the horizon are\footnote{In this section consider only linearized order where the matter stress tensor $T^M$ vanishes, we include it here only as an indicator of how matter couples at next order.}
\be\label{vv} \p_v(D^Ah_{Av}+2M h_{vv})-{1 \over 4M}D^Ah_{Av}-\half D^2h_{vv}=32\pi M^2T^M_{vv}.\ee
\bea \label{va}\p_v(-D_Ah_{vr}-\p_rh_{Av}+{1 \over M}h_{vA}+{1 \over 4M^2}D^Bh_{BA})+D_A\p_r h_{vv}+{1 \over 2M}D_Ah_{vr}\cr+{1 \over 4M^2}D_AD^Bh_{Bv}-{1 \over 4M^2}D^2h_{Av}-{1 \over 4M^2}h_{Av}=16\pi T^M_{Av}.\eea
Using the linear combination of (\ref{vv}) and the divergence of (\ref{va}) given by 
\bea M\p_v(2 h_{vv}+D^2h_{vr}+D^A\p_rh_{Av})-\half D^2h_{vv} -{1 \over 4M}D^AD^B\p_vh_{BA} -MD^2\p_r h_{vv}\cr-{1 \over 2}D^2h_{vr}+{1\over 4M}D^Ah_{Av}=32\pi M^2T^M_{vv}-16\pi MD^AT^M_{Av}\eea
and integrating by parts it finally follows that 
\bea  \label{saz} \Q_f^{\ch}&=&{1\over 8\pi }\int_{\ch}d^2\Theta\sqrt{\g} dvf\bigl( {1 \over 4M}D^AD^B\p_vh_{BA} +32\pi M^2T^M_{vv}-16\pi MD^AT^M_{Av}\cr && ~~~~~~~~~~~+ \half D^2h_{vv}  +MD^2\p_r h_{vv}+{1 \over 2}D^2h_{vr}-{1\over 4M}D^Ah_{Av} \bigr). \eea
We will see that this generates horizon supertranslations after appropriate gauge fixing and boundary conditions. 

\subsection{Gauge fixing and Dirac brackets}

We cannot yet construct Dirac brackets because the presymplectic form $\o$ still has zero eigenvectors given by residual gauge transformations which vanish at $\ch_\pm$ and preserve the Bondi  gauge (\ref{gone})-(\ref{gthree}). We now find the most general such transformation. Differentiating (\ref{gone}) with respect to $r$ and using (\ref{gtwo}) one finds the condition
\be r\p_r^2\zeta^A+2\p_r\zeta^A=0.\ee
The general solution to the above equation is 
\be \zeta^A(r,v,\Theta)=X^A(v,\Theta)+{ 1 \over r}Z^A(v,\Theta).\ee
Substituting $\zeta^A(r,v,\Theta)$ from above into (\ref{gone})-(\ref{gthree}) gives the remaining components of $\zeta^a$, 
\be \label{pazv} \zeta^r=-{r \over 2}D_A\zeta^A,~~~\p_A\zeta^v=\gamma_{AB}Z^{B}.\ee Let us define $\zeta^v=X(v,\Theta)$. Then the most general residual diffeomorphism  $\zeta_X$ for Schwarzschild in Bondi gauge\footnote{This is more general than the usual BMS vector fields discussed e.g. in \cite{bt} as we have not imposed any falloffs.} is parametrized by an arbitrary vector $X^A(v,\Theta)$ 
and an arbitrary scalar $X(v,\Theta)$ on $\ch$ as:
\be \zeta_X=X\p_v-{1 \over 2}(rD_AX^A+D^2X)\p_r+X^A\p_A+{ 1 \over r}D^AX\p_A.\ee
These shift the nonzero metric perturbations to leading order as follows 
\bea \label{resg} \delta   _X h_{vv}&=&{M \over r}D_BX^B+{M \over r^2}D^2X-2V\p_vX-r\p_vD_BX^B-D^2\p_vX,\cr
\delta   _X h_{Av}&=&-{r \over 2}D_AD_BX^B-{1 \over 2}D_AD^2X-V\p_AX+r^2\p_vX_A+r\p_vD_AX,\cr 
\delta   _X h_{AB}&=&r^2(D_AX_B+D_BX_A-\gamma_{AB}D_CX^C)+r(2D_AD_BX - \gamma_{AB}D^2X),\cr
\delta   _X h_{vr}&=& -{1 \over 2} D_BX^B+\p_vX,\eea
where, as usual, $X_A=\g_{AB}X^B$ and $\delta_X$ denotes the Lie action of $\zeta_X$ on Schwarzschild. A supertranslation is 
\be X=f, ~~~X^A=0,\ee
with $\p_vf=0$, while a superrotation is\footnote{Locally imposing the standard Bondi falloff conditions at large $r$ requires $Y^A$ to be locally a conformal Killing vector \cite{bt} and implies $D^2D_AY^A=-2D_AY^A$. We will not impose this restriction herein. The general expression for the charge  is 
$ \Q_{(X,X^A)}=-{1 \over 16 \pi}\int_{\p\Sigma} d^2\Theta \sqrt{\g}r\big[
X^A(r\p_r h_{Av}-{2}h_{Av}+rD^Ah_{vr})-X(D^A\p_r h_{Av} +{2 }h_{vv}  
+(D^2+4V)h_{vr})\big], $
while the central term is 
$ \o(X,X^A; X',X'^A)={1 \over 16 \pi}\int_{\p\Sigma} d^2\Theta \sqrt{\g}r\big[
X(-D^2-2+\frac{6M}{r})D_AX^{\prime A} - (X \rightarrow X^\prime, X^{\prime A}\rightarrow X^A)\big].$} 
\be \label{sr}X={v \over 2}D_AY^A,~~~X^A=Y^A,\ee
with $\p_vY^A=0$. 

In order to invert the symplectic form to get the Dirac bracket we must fix the trivial gauge symmetry, namely those transformations which (unlike supertranslations)  are zero eigenvectors 
of $\o$ and then use the Einstein equation to restrict to the `on-shell' physical phase space. We moreover by hand restrict the phase space via the boundary condition  $\p_v h_{ab}|_{v=\pm\infty}=0$. This excludes the superrotations (\ref{sr}): a looser boundary condition (see \cite{Campiglia:2016jdj,Campiglia:2016hvg,Conde:2016csj,Strominger:2016wns}) is certainly of interest but outside our current scope. Having done so, we will find $\o$ in two steps. First we will construct a reduced $\o^{red}$ in which all gauge freedom is eliminated. We will then add in the non-trivial gauge modes which requires only the boundary expression (\ref{frv}). 
 
 The constraints (\ref{vv}),(\ref{va})  imply that the linear combinations \be \label{eone}\left[{1 \over 4M}D^Ah_{Av}+\half D^2h_{vv}\right]_{\ch_\pm}=0,\ee \be\label{etwo} \left[ D_A\p_r h_{vv}+{1 \over 2M}D_Ah_{vr}+{1 \over 4M^2}D_AD^Bh_{Bv}-{1 \over 4M^2}D^2h_{Av}-{1 \over 4M^2}h_{Av}\right]_{\ch_\pm} =0,\ee
vanish at the horizon boundaries $\ch_\pm$ where we have set $\p_vh_{ab}=0$.  Defining two convenient 
combinations of the metric perturbations $\tilde h$ and $\tilde h_A$ by
\be \tilde h\equiv h_{vr}+2h_{vv}+2M\p_rh_{vv},\ee 
\be \tilde h_A\equiv h_{Av}+2MD_Ah_{vv},   \ee  (\ref{eone}) and (\ref{etwo}) can be rewritten \be D^A \tilde h_A |_{\ch_\pm}=0,\ee 
\be \left[(D^2+1)\tilde h_A-2MD_A \tilde h \right]_{\ch_\pm}=0.  \ee  
These equations have angular momentum $\ell=0,1$ solutions with $\tilde h_A$ a rotational Killing vector and $\tilde h$ a constant on the sphere. These are related to linearized deformations of the angular momentum and mass of the black hole which are not our present interest. While inclusion of these four modes would not change our final conclusions for simplicity we fix them to zero: 
\be \tilde h_A|_{\ch_\pm}=\tilde h |_{\ch_\pm}=0.  \ee  
Under the general residual $X$ and $ X^A$ gauge transformations (\ref{resg}) one has 
\be \delta   _X\tilde h =-(2D^2+1)\p_v X-6M\p_vD_BX^B, \ee
\be \delta   _X\tilde h_A=4M^2(\p_vX_A-\p_vD_AD_BX^B)-2MD_A(D^2-1)\p_v X.\ee
We can use these to set (for $\ell\neq0,1)$)
\be\label{ui} \tilde h=\tilde h_a=0 \ee
everywhere on the horizon $r=2M$. This still leaves unfixed all $\p_vX=\p_vX^A=0$ transformations, which includes supertranslations. 
Using (\ref{ui}) to  eliminate $h_{vr}$ and $h_{Av}$,
 the constraints simplify to 
\be\label{vvv} (D^2-1)\p_v [h_{vv}]=-16\pi MT^M_{vv}=0,\ee
\bea \label{vva}\p_v[2MD_A\p_r h_{vv}-\p_rh_{Av}+{1 \over 4M^2}D^Bh_{BA}]=16\pi T^M_{Av}=0.\eea
The matter sources are set to zero here because we are constructing the Dirac brackets of the linearized theory. We conclude the quantities in square brackets are function of $\Theta$ only.
Using the residual $\p_vX=\p_vX^A=0$ symmetry we may set then to 0:
\be\label{cone} h_{vv}=0,\ee
\be\label{ctwo} \p_rh_{Av}=2MD_A\p_r h_{vv}+{1 \over 4M^2}D^Bh_{BA}.\ee
It can be shown that this completely fixes all the gauge symmetry, including supertranslations. 
Using these relations, the $rr$ and $rv$ components of the Einstein equations reduce to
\be\label{cthree} {1 \over M}\p_rh_{vr}=8\pi T^M_{rr}=0,\ee
\be\label{cfour} -{ 1 \over 2M}(D^2-1)\p_rh_{vv}-{1 \over 16 M^4}D^AD^Bh_{AB}=8\pi T^M_{vr}=0.\ee

The symplectic form (\ref{sform}) involves $ (h_{AB}, h_{vv},\p_rh_{vv}, h_{rv}, \p_rh_{rv}, 
h_{Av},\p_r h_{Av})$. (\ref{ui}) can be used to eliminate $h_{Av}$ and $h_{vr}$ in terms of other variables. (\ref{cone})-(\ref{cfour}) then eliminate $h_{vv},~\p_rh_{Av},~\p_rh_{vr}$ and $\p_rh_{vv}$, expressing everything in terms of the traceless $h_{AB}$.  Denoting such fully gauge-fixed on-shell perturbations by $\bar h_{ab}$, one finds  \bea \label{das} \bar h_{vv}&=&0,\cr \p_r\bar h_{vv}&=&-{1 \over 8M^3}[D^2-1]^{-1}D^AD^B\bar h_{AB},\cr \bar h_{vr}&=&{1 \over 4 M^2}[D^2-1]^{-1}D^AD^B\bar h_{AB},\cr
\bar h_{Av}&=&0,\cr \p_r \bar h_{Av}&=&-{1 \over 4 M^2}D_A[D^2-1]^{-1}D^BD^C\bar h_{BC}+{1 \over 4 M^2}D^B\bar h_{AB},\cr  \p_r \bar h_{vr}&=&0.\eea
It should be noted that $(D^2-1)$ is a negative definite operator and therefore its inverse exists.

A computation  reveals that 
\be \omega^{red}(\bar h,\bar h')={1 \over 64\pi M^2}\int dv d^2\Theta\sqrt{\g}\left[\bar h^{AB}\sigma'_{AB}-\bar h \leftrightarrow \bar h'\right]^{\ch_+}_{\ch_-}.\ee
Here the traceless shear tensor 
\be  \sigma_{AB}=\half\p_vh_{AB} \ee  is the local coordinate-invariant dynamical degree of freedom on the horizon. (\ref{das}) expresses the fact that after complete gauge-fixing and imposition of the constraints all metric components are determined by the shear tensor, up to zero modes. It is of course the point of this paper to carefully understand the zero modes.  

At the level of linearized metric perturbations around Schwarzschild, it appears self consistent to view $\bar h_{ab}$ as a complete set of coordinates on the phase space of the horizon. However as we have seen in section 5, the moment interactions are introduced, pure gauge modes corresponding to supertranslations are excited. The gauge condition (\ref{cone}), which eliminated the rigid supertranslations cannot be enforced. Hence one cannot perturbatively construct the interacting theory beginning from fully gauge fixed modes. One must, at a minimum introduce the supertranslation field 
$ \delta_fg_{ab}$, which we shall see shortly is $not$ a zero mode of the presymplectic form $\o$.\footnote{Other pure gauge modes which are not annihilated by the symplectic form are interesting candidates for further physical degrees of freedom,  but are beyond the scope of the present paper.} 
The fully gauge fixed perturbation $\bar h_{ab}$ is  related to the more general Bondi-gauge perturbation $h_{ab}$ by 
\be h_{ab}=\bar h_{ab}+\delta   _f g_{ab}.\ee
The full symplectic form is 
\be \o^{red}(h,h')=\o(\bar h, \bar h')+\o^{red}(\bar h,\delta   _{f'}g)+\o^{red}(\delta   _fg,\bar h')+\o^{red}(\delta   _fg,\delta   _{f'}g).\ee
The last term is easily seen to vanish, implying there is no classical central term in the  supertranslation algebra.  The middle two terms were essentially computed in (\ref{saz}). Using (\ref{ui}) and setting the sources to zero, one finds \bea  \label{sahz} \o^{red} (h,\delta   _{f'}g)={1\over 16\pi M}\int_{\ch}dvd^2\Theta\sqrt{\g} f'D^AD^B\sigma_{AB}. \eea

 Putting this together and using $h_{AB}=\bar h_{AB}+2M(2D_AD_B f- \gamma_{AB}D^2f)$, one finds 
\bea \omega^{red}( h, h')={1 \over 64 \pi M^2}\int d^2\Theta\sqrt{\g}\int dv  h^{AB}\sigma'_{AB}- h \leftrightarrow h',\eea
This implies the Dirac bracket 
\bea \{ \sigma_{AB}(\Theta,v), h_{CD}(\Theta',v')\}= 32 \pi M^2(\g_{AC}\g_{BD}+\g_{AD}\g_{BC}-
\gamma_{AB}\gamma_{CD})\delta(v-v')\delta^2(\Theta-\Theta'),  
\eea
The expression in parentheses is the deWitt metric \cite{DeWitt:1967yk} for computing distances on
the space of all metrics. Hence
\bea  \Q_f^{\ch}={1\over 16 \pi M}\int_{\ch} dv d^2\Theta\sqrt{\g} fD^AD^B\sigma _{AB} 
. \eea and\bea \{\Q^{\ch}_f,h_{AB}\}&=&2M(2D_AD_Bf-\gamma_{AB}D^2f) \eea
as desired.\footnote{When interactions are included, the charge acquires terms quadratic in $h$, and the right hand side has homogeneous terms such as  
$f\p_vh_{AB}$.}  This equations state that the linearized charge $\Q_f^{\ch}$ is the symplectic partner of the supertranslation zero mode $\delta   _fg_{ab}$. This was guaranteed to work by the general argument of section 7.1 once the physical phase space and sympectic form were properly identified. 

The linearized charge $\Q_f^{\ch}$ is a multipole moment of  the zero mode of the shear tensor
which is the local Cauchy data on the horizon. $\Q_f^{\ch}$ does not vanish for generic shear tensor 
on $\ch$. If it did, it could not generate supertranslations via Dirac brackets on the physical phase space. However, if we look at the space of linearized fluctuations on $\ch$ that can be excited by sending in linearized gravity waves from $\ci^-$, they all have $\Q_f^{\ch}=0$. This follows from the fact that the black hole absorption amplitude for the $\l$th partial wave with frequency $\o$ is proportional to $\o^\l$.\footnote{This has recently been shown to follow directly from supertranslation charge conservation \cite{Avery:2016zce}.} The $\ch$ Cauchy data with nonzero $\Q_f^{\ch}$ would, if evolved backwards with the linearized equations, give perturbations which diverge far from $\ch$ on $\ci^-$.  Nevertheless, we here see that the horizon phase space parametrized by $\bar h_{AB}$ must be enlarged by the symplectic pair $(\Q_f^{\ch},\delta   _fg_{ab})$ in order to have a suitable starting point for the interacting theory in which, as we have seen in section 5, the supertranslation field can not be frozen.  Equivalently, the enlargement of the phase space is required for the existence of  an operator which generates supertranslations everywhere in the spacetime.

\bigskip

\centerline{\bf Acknowledgements}
We are grateful to Abhay Ashtekhar, Glenn Barnich, Piotr Chrusciel, Geoffrey Compere,  Mihalis Dafermos, Sasha Haco, Dan Kapec, Alex Lupsasca, Prahar Mitra, Monica Pate, David Skinner, Bob Wald and S. T. Yau for useful discussions.  This work was supported in part by the Black Hole Initiative, STFC, DOE grant DE-FG02-91ER40654, the George and Cynthia Mitchell Foundation, Trinity College Cambridge, the Templeton Foundation and a Bershadsky Visiting Fellowship to MJP.

\section{Appendix: Some useful formulae}

In this appendix we collect some formulae which we have found useful in our computations. 
In Schwarzschild the nonzero connection coefficients are \bea ~~~\G^A_{rB}={\delta^A_{~B} \over r},~~~\G^v_{AB}=-{r}\gamma_{AB},~~~\G^r_{AB}=-{r}V\gamma_{AB},~~~\G^A_{BC}&=&^{(2)}\G^A_{BC},\cr   \G^r_{vr}=-{M \over r^2} ,~~~\G^v_{vv}={M \over r^2},~~~\G^r_{vv}&=&{M V\over r^2}.\eea
Covariant derivatives on the unit sphere obey
\be [D^2,D_A]X=\p_AX,~~~[D^B,D_A]X_B=X_A.\ee 
One finds at $r=2M$ 
 \bea F_{rv}=\zeta^A(\p_r h_{Av}-{1 \over M}h_{Av})
+\zeta^v(-{1 \over 4M^2}D^Ah_{Av}+{1 \over 4M^2}\p_v\h -{1 \over M}h_{vv}-{1\over 16M^3}\h) \cr  +\p_r \zeta^v h_{vv} +{1 \over 4M^2}
D^A\zeta^vh_{vA}-{1 \over 8M^2} \p_v \zeta^v\h 
+\zeta^r({1 \over M}h_{vr}+{1 \over 8M^3}\h) +{1 \over 8M^2} \p_r \zeta^r\h\eea
On the horizon, supertranslations are given by
\bea\delta   _fh_{vv}&=&{1 \over 4M}D^2f,\cr  
\delta   _f\p_r h_{vv}&=&-{1 \over 4M^2}D^2f,\cr
\delta   _fh_{Av}&=&-{1 \over 2}\p_AD^2f,\cr \delta   _f\p_r h_{Av}&=&-{1 \over 2M}\p_Af,\cr
\delta   _fh_{AB}&=&{2M}(2D_AD_Bf- \gamma_{AB}D^2f),\cr 
D^2D_Af&=&D_AD^2f+\p_Af \eea
From the linearized constraints on the horizon (\ref{vv})(\ref{va})
\be \p_rh_{vv}=-{1 \over 4M}h,   ~~~h_{vr}={1 \over 2}h,~~~\p_rh_{Av}=-{1 \over 2}D_A h+{1 \over 4 M^2}D^Bh_{AB}, ~~~h_{vv}=h_{Av}=0.\ee
\bea \nabla^bh_{bv}
&=&\half \p_vh \cr
\nabla_b h_{vc}-\nabla_v h_{bc} &=&\p_b h_{vc}-\p_vh_{bc}-\G^a_{bc}h_{va}+\G^a_{vc}h_{ba} \cr \nabla_rh_{vv}-\nabla_v h_{rv}
&=&-{1 \over 2 }\p_vh ,\cr  \nabla_B h_{vC}-\nabla_v h_{BC}&=&\-\p_vh_{BC}.  
\eea
Then the pre-symplectic current becomes
\bea J_v&=& {\sqrt{-g} \over 16\pi}[{1 \over 4}h\p_vh'-\half h\p_vh'+{1 \over 4}h\p_vh'  {1 \over 2} h^{AB}\p_vh'_{AB}-h \leftrightarrow h' ]\cr &=&{\sqrt{-g} \over 32\pi} h^{AB}\p_vh'_{AB}-h \leftrightarrow h' \eea

\end{document}